\documentclass{ws-procs9x6}

\begin{document}

\title{Superstring Relics, Supersymmetric Fragmentation and UHECR}

\author{ A. Cafarella, C. Corian\`{o}\footnote{
Plenary Talk presented by Claudio Corian\`{o} 
at the First Intl. Conf. on String Phenomenology, 
Oxford, UK, July 6-11 2002},
  M. Guzzi and  D. Martello}
\address{Dipartimento di Fisica\\ 
Universita' di Lecce \\
INFN Sezione di Lecce\\
Via Arnesano 73100, Lecce, Italy \\ }

\maketitle

\abstracts{ 
Superstring theory predicts the existence of relic metastable particles 
whose average lifetime is longer than the age of the universe and which 
could, in principle, be good dark matter candidates. 
At the same time, these states would be responsible for the  Ultra High 
Energy Cosmic Rays (UHECR) events which will be searched for by 
various experimental collaborations in the near future. 
We describe a possible phenomenological path which could be followed in 
order to search for new physics in their detection.}

\section{Introduction}
Strings are with no doubt the richest and most fascinating theoretical 
constructs of the last decades, but, at this time, no experimental 
signal has emerged pointing toward a validation/ falsification of the theory.
This is the main motivation for studying string model building \cite{af}.

On the other hand, the success of the Standard Model has its own unescapable 
limitations, since the origin of the mass in the model is still 
under question and issues of naturalness indicate that new physics might 
be at the door at forthcoming collider experiments. In the meanwhile, we have 
also seen that important new results  
from the astroparticle side -such as neutrino oscillations- 
are getting more and more entangled with 
``traditional'' collider studies to give us a broader perspective 
of what may lay beyond  the high energy frontier. 

\section{Cosmic Rays}
Particle Astrophysics, in particular cosmic rays astrophysics, 
may not be able to reach the level of accuracy that we expect 
from other kinds of experiments, such as collider experiments, but the 
technology is rapidly changing and we may hope for serious improvements 
within a decade or even less.   

It is to be remarked that theoretical methods and tools of analysis 
which are common in collider phenomenology are not immediately applicable 
to astroparticle physics, since the scales entering in the dynamics, 
for instance in the UHECR dynamics \cite{SubirSarkar}, are in practice much larger that those we are used to in collider studies. Neverthless, it is still interesting to see 
what we can say about processes at the Greisen-Zatsepin-Kuzmin (GZK) 
($\sim 10^{19}$ eV) cutoff in terms of standard QCD tools based on perturbation theory.  

Recently, it has been shown that the renormalization group can be used 
 to describe the dynamics of the UHECR is the so called 
`` top-down'' approach to their origin. In particular, the impact of supersymmetry in the 
re-organization of the spectrum of heavy relics fragmenting into 
ordinary hadrons has been quantified \cite{coriano-faraggi}. 
The re-adjustement of the 
spectrum obtained is, of course, dependent on the type of evolution model used 
to generate supersymmetric fragmentation functions from the high scale, 
down to the low energy scale where ordinary hadronization takes place. 
In \cite{coriano-faraggi} the model analized is based on the 
mixing of supersymmetric anomalous dimensions with the corresponding 
QCD ones with vanishing initial conditions at the susy threshold(s) \cite{CC}.

We have to mention that the analysis presented in \cite{coriano-faraggi} 
can be also applied to other mechanisms held responsible for the origin 
of the UHECR, such as Z-bursts and others, since it is the 
{\em first impact} (the interaction of the protons of the primary 
ray with the atmosphere)  to affect the evolution of the cascade and 
to arrange the multiplicity distributions and the associated shape variables. 
We recall that for GZK energies, the center of mass energy of the incoming nucleon with the nucleon in the atmosphere is of the order of several hundreds of TeV's.  

\section{Wilson Lines Breaking and Fractionally Charged States} 
One of the solution to the problem of the origin of the UHECR, 
as we have mentioned above, comes from the decay of long-lived super-heavy states whose mass is in the $10^{12-15}$ GeV range. 
The primaries, according to this hypothesis, would be generated 
in the galactic halo, and the GZK cutoff would not apply. 
The mass required for the meta-stable state, whose lifetime is between 
$10^{17}-10^{27}$ sec, is about $10^{12-13}$ GeV. Their abundancy 
($\sim 5\times 10^{-11}$) is constrained by the observed flux of the UHECR events. 

Superstring theory can naturally account for the meta-stable states, through a 
stabilization mechanism due to the breaking of the non-abelian gauge symmetries by Wilson lines. This mechanism \cite{CCF} gives rise 
to states in the string spectrum which carry standard charge under the 
Standard Model, but fractional charge under an additional
$U(1)_{Z^\prime}$ gauge group. The existence of fractionally 
charged states can be regarded as a standard consequence 
of string unification. 

\section{ Detecting UHECR and the Case for New Physics}
The Auger experiment is constructing two $300$ Km$^2$ grids of detectors spaced at 1.5 km intervals, one in the northern hemisphere and a second one in the 
southern hemisphere. The possibility 
to detect fluorescence radiation along the slanted 
path of the incoming primary has also been taken into account and implemented 
through  flye's eye detector arrays, beside the usual Cerenkov detectors, laid on the ground.

In the top-down approach to the formation of UHECR

1) a relic decays and generates a given initial distribution of primaries; 

2) primary particles propagate and mostly protons survive; 

3) primary protons collide with the atmosphere thereby generating 
atmospheric showers; 

4) multiplicity of distributions and their deposited energy are detected on the ground. 

If we aim at unveiling new physics from UHECR experiments, we should 

1) allow in the analysis of the first impact of the primary the  possibility 
of new interactions;

2) allow a modification of the fragmentation region due to the new 
interactions 
(from angular ordering to small $p_T$ exchanges, now with susy).

In standard cosmic rays applications, the simulations 
of cosmic rays events are performed using monte carlo event generators 
whose purpose is 

1) to generate a first collision with the nuclei in the atmosphere  and 

2) track down each secondary ray produced to some 
selected observation levels. 

The type of interaction processes implemented in the existing 
simulation programs do not, at the moment, include any new physics yet.

\section{New Models of Interaction and their Simulations}
It is possible to introduce new models of interactions, 
(in our case, specifically, supersymmetric models) and interface them with the traditional monte carlo 
generators which generate the profile of the final shower and see the 
enhanced/modified effects on the multiplicities at various 
observation levels in the atmosphere. This study is on the way and results will 
be presented elsewhere. However, to illustrate the shape 
of the distribution of multiplicites that one expects from the simulations 
of this type, we have included 3 Figs., 
which we have generated using the simulation program CORSIKA \cite{corsika}, 
interfaced with the program SYBILL \cite{sybill} for the generation of the hadronic interactions. The Figs. show separate multiplicities for various particles plotted 
against rapidities and/or radial distances, the latter measured respect to 
an axis running perpendicular to the plane of the detector. 
The geometry is simplified here for a zero zenith angle of impact of 
the ray, as measured by an observer on the ground of the detector. 

It is to be 
remarked that the experiments will be measuring only the energy 
deposited and other measurements, such as of rapidities or $p_T$ 
distributions on the final showers will not be possible at any of these 
experiments. Neutrinos are also not detected and 
therefore these plots are mostly of theoretical interests. 

As one can see, multiplicity distributions of the 3 main particles 
are clearly separated, the photon showing higher statistics compared to 
the other particles. In the simulations we have kept the depth of the 
first impact to be random, at an energy below the GZK cutoff ($10^5$ TeV). 
We remark that current hadronic interaction models may 
not be appropriate at the GZK cutoff. 
The results show a fast increase in the number of the secondaries as we raise the 
collision energy. To allow comparisons, the multiplicities 
have been normalized in such a way that the area under each distribution 
is set to be one. 

\section{Astro QCD/ SUSY-QCD: Perspectives}
Among the issues that need to be analized in the context of Ultra High 
Energy QCD (Astro-QCD) is the role played by supersymmetry 
in the fragmentation, the role of mini-jet cross sections and of vacuum 
exchanges (with and without supersymmetry) near the cutoff, where 
theory is still lagging compared to the planned experiments. 

Numerical solutions of the 
Renormalization Group Equations \cite{CC} 
have been obtained recently and indicate that 
- at least at large momentum transfers- the impact of supersymmetry on 
the evolution of the cascades is small and comparable to other hadronic
background \cite{coriano-faraggi}.  
To support this point, in Fig.4 \cite{coriano-faraggi} we 
show results obtained by solving the Renormalization Group Equations 
for the following inclusive observables,

\begin{equation}
R_{QCD}^h(Q^2)=\sum _{i=1}^{n_f}e_i^2\int_{z_{min}^h}^1
d\,z\left(D_{q_i}^h(z,Q^2)
+ D_{\bar{q}_i}^h(z,Q^2)\right)
\end{equation}
and

\begin{equation}
R_{SQCD}^h(Q^2)=\sum _{i=1}^{n_f}e_i^2\int_{z_{min}^h}^1
d\,z\left(D_{q_i}^h(z,Q^2) + D_{\bar{q}_i}^h(z,Q^2)
+ D_{\tilde{q}_i}^h(z,Q^2) + D_{\bar{\tilde{q}}_i}^h(z,Q^2)\right)
\nonumber\\
\end{equation}
 with $D_{q_i}^h(z,Q^2)$ being the fragmentation of function of quark 
flavour $i$ into hadrons $h$,
and with $ z_{min}^h= m_h/(Q/2)$ being the minimum fractional energy required for
the
fragmentation to take place. $e_i$ are the charges of the $n_f$ quark and
squark flavours. This result has been obtained solving approximately 100 
coupled equations of DGLAP-type, as described in \cite{coriano-faraggi}. 
Fig.~4 describes the fragmentation into final protons and compares 
the QCD case (curve above) to the Susy QCD (SQCD) one (curve below). 
 Again, we see that there are indications that susy effects, 
linked to the mixing of supersymmetric operators to QCD light-cone operators, 
are at the few percent level. These differences, although small, 
can't be discarded, since in monte carlo simulations of showers 
these effects change sensitively the hadronization process.
If they combine to give variations in the total cross sections 
at the 10 percent level, they may affect  
the profile of the showers \cite{AC}. 

 Although these results are encouraging, 
and theorists have shown a great deal of interest on the matter 
\cite{other}, more work in this direction, however, is needed.

\centerline{Acknowledgement}
We thank Alon Faraggi for collaboration on related work 
and Subir Sarkar for discussions. The work of A.C., C.C. and M.G. 
is partly supported by INFN (iniziativa specifica BA21).

\begin{figure}[th]
\centerline{\includegraphics[angle=-90,width=.8\textwidth]{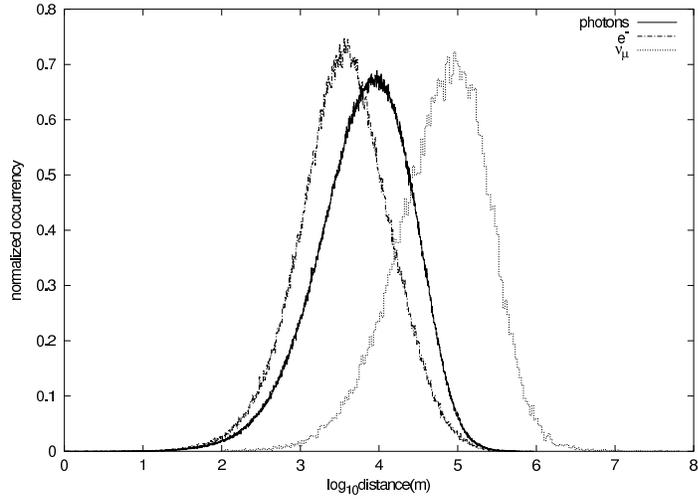}}
\caption{ Plot pf the normalized occurrencies of 
$\gamma$, $\nu_\mu$ and $e^-$ assuming an impact energy of $10^5$ TeV as a function of distance on a logarithmic scale\label{inter1}}
\end{figure}

\begin{figure}[th]
\centerline{\includegraphics[angle=-90,width=.8\textwidth]{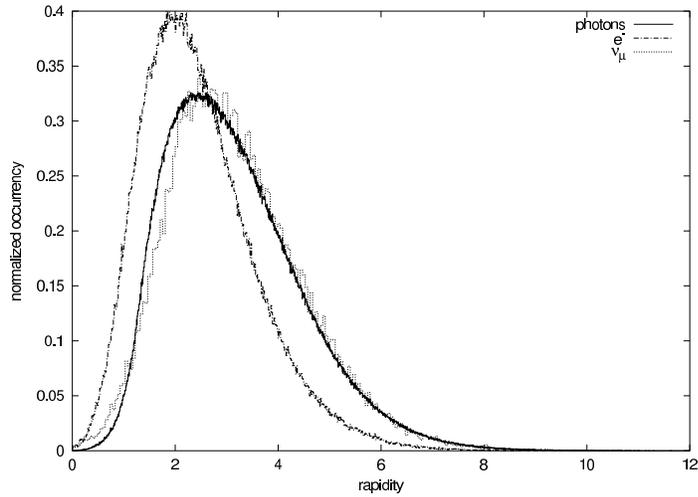}}
\caption{distributions of rapidities of $\gamma$, $\nu_\mu$ and $e^-$ 
at the same energy as in Fig.~1.\label{inter2}}
\end{figure}

\begin{figure}[th]
\centerline{\includegraphics[angle=-90,width=.8\textwidth]{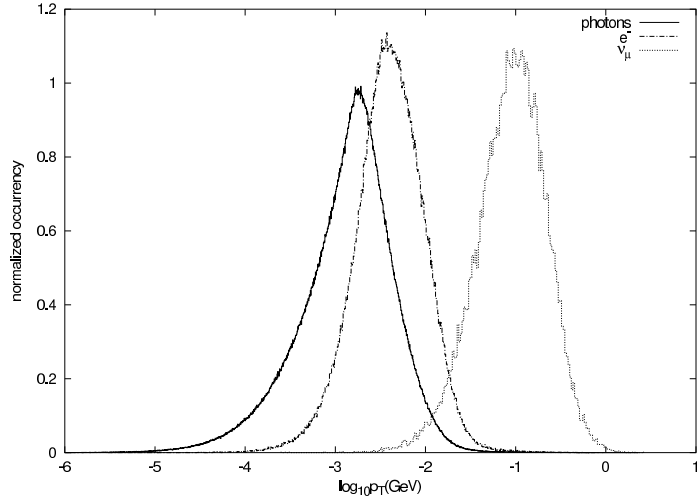}}
\caption{ Plot pf the normalized occurrencies of 
$\gamma$, $\nu_\mu$ and $e^-$ assuming an impact energy of $10^5$ TeV as a function of transverse momentum $p_T$\label{inter3}}
\end{figure}

\begin{figure}[th]
\centerline{\includegraphics[angle=0,width=.6\textwidth]{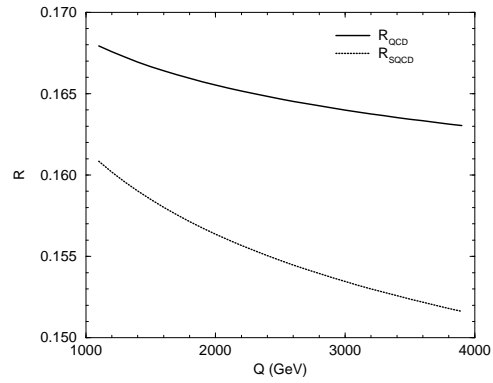}}
\caption{ Plot of the R-function describing the fragmentation of quarks into 
final protons with supersmmetry (curve above) and without supersymmetry 
(curve below) versus Q, the initial fragmentation energy.}

\end{figure}

\end{document}